\documentclass[11pt]{article}
\usepackage[utf8]{inputenc}
\usepackage[left=1in, right=1in, bottom=1.15in, top=1.1in]{geometry}

\usepackage{MG}

\title{A Decomposition Theorem and a Simple Algorithm \\for Contraction Fixed Points}

\author{
Xi Chen\\Columbia University\\\url{xichen@cs.columbia.edu}\hspace{-0.2cm}
\and Yuhao Li\\Columbia University\\\url{yuhaoli@cs.columbia.edu}\hspace{-0.2cm}
\and Mihalis Yannakakis\\Columbia University\\\url{mihalis@cs.columbia.edu}\vspace{0.1cm}
}

\date{}
\begin{document}
\maketitle
\begin{abstract}

We prove a decomposition theorem for the problem of computing an approximate fixed point of a contraction map $f\colon [0,1]^k \to [0,1]^k$ under the $\ell_{\infty}$-norm. 

After a preliminary version of the decomposition theorem appeared on arXiv~\cite{CLY26}, Feodorov and Haslebacher~\cite{FH26}  gave a striking application of the theorem, showing that it can be used to convert any $\operatorname{poly}(k,\log(1/\varepsilon))$-query algorithm with running time $(k\log(1/\varepsilon))^{O(k)}$ into an algorithm with running time $(k\log(1/\varepsilon))^{O(\sqrt{k})}$. They instantiated this framework with an $O(k^2\log(1/\varepsilon))$-query algorithm with the desired running time, thereby obtaining the first subexponential-time algorithm for contraction fixed points under the $\ell_{\infty}$-norm and, via well-known reductions, the first deterministic subexponential-time algorithm for simple stochastic games and mean-payoff games. 

In addition to the decomposition theorem, as our second main result in this paper, we show that the
$O(k\log(1/\varepsilon))$-query algorithm of~\cite{CLY25} can be implemented in time $(k\log(1/\varepsilon))^{O(k)}$; our algorithm may be viewed as conceptually simpler compared to the algorithm of \cite{FH26}. A key observation here is that the function constructed in~\cite{CLY25}, whose fixed points are used as the next query points, turns out to be nonexpansive under the $\ell_{\infty}$-norm, so the brute-force procedure can be replaced by known algorithms for contraction fixed points.

\end{abstract}

\thispagestyle{empty}
\newpage 
\setcounter{page}{1}
\section{Introduction}

A map $f:\calM\mapsto \calM$ defined on a complete metric space $(\calM,d)$ is called a $(1-\gamma)$-\emph{contraction} map for~some $\gamma\in(0,1]$ if it satisfies 
$d(f(x),f(y))\leq (1-\gamma)\cdot d(x,y)$ for all $x,y\in\calM$. Banach~\cite{Ban1922} proved~in 1922 that every contraction has a \emph{unique} fixed point. 
Notably, his proof is constructive: iteratively applying the contraction map $f$ starting from any point $x\in \calM$ always converges to its unique fixed point.
Banach's foundational result, now known as the Banach fixed point theorem, has profoundly influenced numerous disciplines over the past century, including mathematics~\cite{CL55,nash1956imbedding,gunther1989}, economics and game theory~\cite{Shapley53,stokey1989recursive}, and computer science~\cite{bellman1957markovian,howard1960dynamic,De67}. From an algorithmic perspective, many~fundamental computational problems can be reduced to finding the unique fixed point or a high-precision approximate fixed point, which motivates the study~of~fixed point computation in contractions \cite{STW93,HKS99,SS02,SS03,DP11,DTZ18,FGMS20,CLY25,batzioumonotonecontraction,haslebacher2025query,CLY26,FH26}.



One of the most natural and important metric spaces to study 
the computation of \mbox{contraction} fixed points 
 is the cube $[0,1]^k$ with respect to the $\ell_{\infty}$-norm. 
In particular, {well-studied} games~from formal verification, such as parity games~\cite{EJ91,CJKLS22}, mean-payoff games~\cite{ehrenfeucht1979positional,ZP96}, as~well as simple stochastic games~\cite{Con92}, can all be reduced to finding~an $\eps$-fixed point\footnote{A point $x\in \calM$ is said to be an $\eps$-approximate fixed point, or simply an $\eps$-fixed point, of $f$ if $d(x,f(x))\le \eps$. For $[0,1]^k$ with respect to the $\ell_{\infty}$-norm, this means an $x\in [0,1]^k$ that satisfies $\|f(x)-x\|_{\infty}\le \eps$.} of~a $(1-\gamma)$-contraction $f:[0,1]^k\mapsto [0,1]^k$ under the $\ell_{\infty}$-norm, where the dimension $k$ is polynomial but $\eps$ and $\gamma$ are both inverse exponential in the input size of the game. These games are not only important in verification, but also have an intriguing complexity status: their decision versions are in $\UP\cap\coUP$ but whether polynomial-time algorithms exist remains long-standing open questions. Indeed, the simplest proof of their $\UP\cap\coUP$-membership is based on the contraction fixed point and crucially utilizes the uniqueness guaranteed by Banach's  theorem~\cite{jurdzinski1998deciding,EY10}. 

In this paper, we work on 
  the well-studied \emph{black-box} model for contraction fixed points over $[0,1]^k$ with respect to the $\ell_{\infty}$-norm. 
The problem, denoted by $\Contraction_\infty(\eps,\gamma,k)$, is as follows.
An algorithm is given query access to an unknown $(1-\gamma)$-contraction map $f:[0,1]^k\mapsto[0,1]^k$: in each round, the algorithm can send a point $x\in[0,1]^k$ to the oracle and receive $f(x)$ in response. The goal of the algorithm is to find an $\eps$-fixed point of $f$. 
We distinguish two complexity measures in this model: query complexity and time complexity.
The \emph{query complexity} of an algorithm is the number of oracle queries it uses, and the \emph{time complexity} is the total computational time, assuming that the oracle answers each query in one time step, both in the worst case.
By definition, the time complexity is always at least as large as the query complexity, but not necessarily vice versa.\medskip

\noindent \textbf{Prior work on {\normalfont$\Contraction_{\infty}(\eps,\gamma,k)$}.} While Banach's value iteration method works for any metric space $(\calM,d)$, it requires $O((1/\gamma)\log(1/\eps))$ time to converge to an $\eps$-fixed~point. 

The first algorithms designed specifically for $\Contraction_{\infty}(\eps,\gamma,k)$ over $[0,1]^k$  with respect~to the $\ell_{\infty}$-norm were by Shellman and Sikorski. They first gave an $O(\log (1/\eps))$-time algorithm when $k=2$~\cite{SS02}. Later on, they provided an $O(\log^k(1/\eps))$-time algorithm for general $k$~\cite{SS03}, which remained the best algorithm when $k$ is smaller than $\log(1/\eps)$ and $\log(1/\gamma)$ until very recently.

Recently  
Chen, Li and Yannakakis~\cite{CLY25}  gave an algorithm for $\Contraction_{\infty}(\eps,\gamma,k)$ with query complexity $O(k\log(1/\eps))$. In terms of time complexity, as reviewed later in \Cref{sec:review},  their algorithm uses brute force to find the next query point, which may take up to $(1/\eps)^k$ time. 
 
We notice that all these upper bounds 
do not \mbox{depend} on $\gamma$; indeed, algorithms of \cite{SS02,SS03,CLY25} can solve the more general problems of $\NonExp_{\infty}(\eps,k)$, the problem of finding an $\eps$-fixed point in a nonexpansive\footnote{A map $f:\calM\rightarrow \calM$ is said to be nonexpansive if $d(f(x),f(y))\le d(x,y)$ for all $x,y\in\calM$.} map $f: [0,1]^k\rightarrow [0,1]^k$ with respect to the $\ell_{\infty}$-norm. 
As discussed below, this will remain the case for all the results presented in this paper.\medskip

\noindent\textbf{Our first contribution.}
Our first result is a dimension-based decomposition theorem for $\eps$-fixed points of nonexpansive maps under the $\ell_{\infty}$-norm.

\begin{theorem}\label{thm: linfty decomposition}
	Suppose that $\NonExp_{\infty}(\eps,a)$ can be solved in $q(\eps,a)$ queries and $t(\eps,a)$ time, and $\NonExp_{\infty}(\eps,b)$ can be solved in $q(\eps,b)$ queries and $t(\eps,b)$ time. Then $\NonExp_{\infty}(\eps,a+b)$ can be solved with 
    query complexity $q(\eps,a)\cdot q(\eps,b)$  and time complexity $$O\Big (\big(t(\eps,a)+q(\eps,b)\big)\cdot q(\eps,b) + t(\eps,b)\Big).$$ 
\end{theorem}
The roles of $a$ and $b$ can be interchanged; we state only one orientation for simplicity.
Combining it with the $O(\log(1/\eps))$-time algorithm for $\NonExp_{\infty}(\eps,2)$ of \cite{SS02}, our decompo\-sition theorem immediately yields an $O(\log^{\lceil k/2\rceil}(1/\eps))$-time algorithm for general $k$, thus achieving a quadratic speedup over the previous best running time bound of~\cite{SS03}.
As it turns out, however, our decomposition theorem enables a far more substantial speedup.

After a preliminary version of the decomposition theorem appeared on arXiv~\cite{CLY26}, Feodorov and Haslebacher~\cite{FH26} gave a striking application of the theorem, showing that  
\begin{enumerate}
\item[] \emph{Any algorithm for $\NonExp_\infty(\eps,k)$ with  $\operatorname{poly}(k,\log(1/\varepsilon))$ queries and $\smash{(k\log(1/\varepsilon))^{O(k)}}$    time\vspace{0.015cm}\\ can be converted into a 
$\smash{(k\log(1/\varepsilon))^{O(\sqrt{k})}}$-time algorithm for $\NonExp_\infty(\eps,k)$.}
\end{enumerate}
Indeed, building on the query algorithm of~\cite{CLY25}, they gave an algorithm with $O(k^2\log(1/\varepsilon))$ queries and $\smash{(k\log(1/\varepsilon))^{O(k)}}$ time, thereby~obtaining the first subexponential $\smash{(k\log(1/\varepsilon))^{O(\sqrt{k})}}$-time algorithm for $\NonExp_{\infty}(\eps,k)$. 
This  leads~to
the first~deterministic, $\smash{|\calG|^{O(\sqrt{n})}}$-time algorithm for simple stochastic games and mean-payoff games, where $n$ denotes the number of vertices in the graph and $\smash{|\calG|}$ denotes the input size of the game.


We note that analogous decomposition theorems have been established for Tarski fixed points \cite{FPS22,CL22} as well as monotone contractions~\cite{batzioumonotonecontraction}. The precise statements and underlying proof techniques of these theorems, however, differ substantially from ours.\medskip

\noindent\textbf{Our second contribution.}
As our second contribution in this paper, we present a new algorithm for $\NonExp_{\infty}(\eps,k)$ with query complexity $O(k\log(1/\eps))$ and time complexity $(k\log(1/\eps))^{O(k)}$. 

\begin{restatable}{theorem}{theoremnontrivialimplementation}\label{theorem: non-trivial implementation}
    There is an algorithm for $\NonExp_{\infty}(\eps,k)$ with query complexity $O(k\log(1/\eps))$ and time complexity $(k\log(1/\eps))^{O(k)}$. 
\end{restatable}

As discussed above, our algorithm can also be plugged into the framework of \cite{FH26} to yield a $\smash{(k\log(1/\eps))^{O(\sqrt{k})}}$-time algorithm for $\NonExp_\infty(\eps,k)$.
Despite achieving the same time complexity, 
 our algorithm may be viewed as conceptually simpler, as we explain below.

As reviewed in \Cref{sec:review}, the key technical ingredient behind the $O(k\log(1/\eps))$-query upper bound of  \cite{CLY25} is the existence of a so-called \emph{balanced point} for any measurable set $\smash{S\subseteq [0,1]^k}$. This was established by designing a continuous map $g_S:[0,1]^k\rightarrow [0,1]^k$; by Brouwer's fixed point theorem, $g_S$ has a fixed point, which is shown to be a balanced point of $S$.

The key observation behind our algorithm for \Cref{theorem: non-trivial implementation} is that the map $g_S$ designed in~\cite{CLY25} is  not only continuous, but actually nonexpansive with respect to the $\ell_{\infty}$-norm! As a result, one can find an approximate fixed point of $g_S$ using the algorithm of Shellman and Sikorski~\cite{SS03} as a subroutine, which can be shown to be an approximate balanced point of $S$.
Replacing balanced points in \cite{CLY25} with approximate balanced points leads to an algorithm with the desired query and time complexities for $\NonExp_\infty(\eps,k)$ given in \Cref{theorem: non-trivial implementation}. See more details in \Cref{section: implementation}.

\section{Preliminaries}
We start with the definitions of contraction and nonexpansive maps.

\begin{definition}[Contraction]
Let $0<\gamma<1$ and $(\calM,d)$ be a metric space. A map $f:\calM\mapsto\calM$ is a $(1-\gamma)$-\emph{contraction map} with respect to $(\calM,d)$ if $d(f(x),f(y))\leq (1-\gamma)\cdot d(x,y)$ for all $x,y\in \calM$.

A map $f:\calM\mapsto\calM$ is said to be \emph{nonexpansive} if 
  $d(f(x),f(y))\leq d(x,y)$ for all $x,y\in \calM$.
\end{definition}

We will focus on $[0,1]^k$ with respect to the
  $\ell_\infty$-norm in this paper. 
Every contraction map~$f$~has a unique fixed point, i.e., $x^*$ with $f(x^*)=x^*$ by Banach's fixed point theorem. Every nonexpansive map,
on the other hand, has a (not necessarily unique) fixed point by Brouwer's fixed point theorem.
We study the following 
  contraction fixed point problem:

\begin{definition}[$\Contraction_{\infty}(\eps,\gamma,k)$]
In problem $\Contraction_{\infty}(\eps,\gamma,k)$, we are given 
  oracle access to a $(1-\gamma)$-contraction map $f$ over $[0,1]^k$ with respect to the $\ell_{\infty}$-norm, i.e., $f$ satisfies
$$
\norm{f(x)-f(y)}_{\infty}\le (1-\gamma)\cdot \norm{x-y}_{\infty},\quad \text{for all $x,y\in [0,1]^k$}.
$$
The goal is to find an $\eps$-fixed point of $f$, i.e., a point $x \in [0,1]^k$ such that $\norm{f(x )-x}_{\infty}\leq \eps$. 
\end{definition}

We define similar problems for nonexpansive maps:

\begin{definition}[$\NonExp_{\infty}(\eps,k)$]\label{def:nonexpansive}
In problem $\NonExp_{\infty}(\eps,k)$, we are given 
  oracle access to a~nonexpansive map $f :[0,1]^k\rightarrow [0,1]^k$ with respect to the $\ell_{\infty}$-norm, i.e., $f$ satisfies
$$
\norm{f(x)-f(y)}_{\infty}\le \norm{x-y}_{\infty},\quad \text{for all $x,y\in [0,1]^k$}.
$$
The goal is to find an $\eps$-fixed point of $f$, i.e., a point $x\in [0,1]^k$ such that $\|f(x)-x\|_{\infty}\le \eps$. 
\end{definition}

We will use the following notation. 
Given a positive integer $m$, we write $[m]$ to denote the set $\{1,\ldots,m\}$. Given $\alpha,\beta\in\mathbb{R}^m$, we say $\alpha\preceq \beta$ if $\alpha_i\leq \beta_i$ for all $i\in[m]$. Given $\alpha,\beta\in\mathbb{R}^m$ such that $\alpha\preceq \beta$ and $\eps>0$, we let
$\calB(\alpha,\beta)\coloneqq \set{x: \alpha\preceq x \preceq \beta}$ denote the \emph{box}  defined by $\alpha$ and $\beta$, and let 
$$\calB(\alpha,\beta,\pm \eps)\coloneqq \big\{x\in \mathbb{R}^m: (\alpha_1-\eps,\cdots,\alpha_m-\eps)\preceq x \preceq (\beta_1+\eps,\cdots,\beta_m+\eps)\big\}.$$

Our decomposition theorem, while stated for $\NonExp_\infty(\eps,k)$ in \Cref{thm: linfty decomposition}, will be established in \Cref{section: decomposition} 
for a variant of $\NonExp_{\infty}(\eps,k)$, denoted by $\NonExp_{\infty}^\dagger(\eps,k)$ and defined below.~We remark that reductions between $\NonExp_{\infty}(\eps,k)$ and $\NonExp_{\infty}^\dagger(\eps,k)$ are known for both directions, which imply that they have  asymptotically the same query and time complexities~\cite{CLY26,FH26}. Nevertheless, for technical reasons, we will prove our decomposition theorem   for $\NonExp_{\infty}^\dagger(\eps,k)$. We use this expanded-range formulation because the restricted slice map need not map the restricted box into itself; its image is only guaranteed to lie within the box expanded by $\eps$ in each coordinate.

\begin{definition}[$\NonExp_{\infty}^\dagger(\eps,k)$]\label{def:nonexpansivegeneralization}
In $\NonExp_{\infty}^\dagger(\eps,k)$ we are given $\alpha_{\min},\alpha_{\max}\in [0,1]^k$~with $\alpha_{\min}\preceq \alpha_{\max}$ and oracle access to a nonexpansive map $f:\calB(\alpha_{\min},\alpha_{\max})\rightarrow \calB(\alpha_{\min},\alpha_{\max},\pm \eps)$
with respect to the $\ell_{\infty}$-norm. The goal is to find an $\eps$-fixed point of $f$, i.e., a point $x\in \calB(\alpha_{\min},\alpha_{\max})$ such~that $\|f(x)-x\|_{\infty}\le \eps$.

\end{definition}
Note that a solution to $\NonExp_{\infty}^\dagger(\eps,k)$ always exists. To see this, we write  $g:\calB(\alpha_{\min},\alpha_{\max})\rightarrow \calB(\alpha_{\min},\alpha_{\max})$ to denote the map obtained by truncating $f$ as follows:
$$
g(x)_i:=\max\big(\min(f(x)_i,\alpha_{\max,i}),\alpha_{\min,i}\big),\quad\text{for every $i\in [k]$.}
$$
It is easy to verify that $g$ remains nonexpansive and thus, has a fixed point $x$ by Brouwer's fixed point theorem.
Any fixed point of $g$ must be an $\eps$-fixed point of $f$.

\section{The Decomposition Theorem}\label{section: decomposition}

In this section, we prove the decomposition theorem, restated below.

\newtheorem*{theoremrepeat}{\Cref{thm: linfty decomposition}}

\begin{theoremrepeat}[Formal Version]
Suppose that $\NonExp^\dagger_{\infty}(\eps,a)$ can be solved in $q(\eps,a)$ queries and $t(\eps,a)$ time, and $\NonExp^\dagger_{\infty}(\eps,b)$ can be solved in $q(\eps,b)$ queries and $t(\eps,b)$ time. Then $\NonExp^\dagger_{\infty}(\eps,a+b)$ can be solved with query complexity $q(\eps,a)\cdot q(\eps,b)$  and time complexity $$O\Big (\big(t(\eps,a)+q(\eps,b)\big)\cdot q(\eps,b) + t(\eps,b)\Big).$$
\end{theoremrepeat}

\noindent\textit{Remark.} Before proving \Cref{thm: linfty decomposition}, we note that an earlier version of this work~\cite{CLY26} stated a weaker running-time bound $O((t(\eps,a)+q(\eps,b))\cdot t(\eps,b)).$ The same algorithm and analysis there in fact yield the stronger bound stated above, as noted by Feodorov and Haslebacher in~\cite{FH26}. 


Before the proof, we would like to provide some intuition on how a decomposition theorem works and what are the main obstracles we need to overcome. At a high level, our decomposition combines solvers in dimensions $a$ and $b$ to solve the $(a+b)$-dimensional problem. For each query $q^t$ of the $b$-dimensional outer solver $\calA_2$, we fix the last $b$ coordinates to $q^t$, use $\calA_1$ to find an $\eps$-fixed point $y^{(t,*)}$ of the resulting $a$-dimensional slice, and return the last $b$ coordinates of $f(y^{(t,*)},q^t)$ as the answer $v^t$. The main difficulty is that independently chosen slice solutions need not yield answers consistent with a nonexpansive outer map, as required by $\calA_2$. Since this map is defined on the fly, we must maintain consistency online as each new query arrives. We do so by requiring each $y^{(t,*)}$ to lie within $\ell_\infty$-distance $\norm{q^t-q^r}_\infty$ of every previous $y^{(r,*)}$. Here the geometry of $\ell_\infty$ is crucial: these constraints define axis-aligned boxes, and consistency of previous choices guarantees that these boxes and the original domain intersect pairwise, which implies a common intersection. This intersection is a nonempty box $\calB(p^t_{\min},p^t_{\max})$ that the slice maps into its $\eps$-expansion, so we can run $\calA_1$ on the restricted slice. Moreover, distances across coordinate blocks combine by taking a maximum, so the bound on the inner solutions and nonexpansiveness of $f$ ensure $\norm{v^t-v^r}_\infty\leq\norm{q^t-q^r}_\infty$. When $\calA_2$ finds an $\eps$-fixed point, the same maximum rule ensures that combining it with its slice solution yields an $\eps$-fixed point of $f$.

\begin{algorithm}[!t]
	\caption{Algorithm for $\NonExp^\dagger_{\infty}(\eps,{a+b} )$ via $\NonExp^\dagger_{\infty}(\eps,a)$ and 
	  $\NonExp^\dagger_{\infty}(\eps,b)$}
	\label{alg: linfty}
	\KwIn{An $\eps\in(0,1]$, $\alpha_{\min},\alpha_{\max}\in[0,1]^a$, $\beta_{\min},\beta_{\max}\in[0,1]^b$ such that $\alpha_{\min}\preceq \alpha_{\max}$ and $\beta_{\min}\preceq \beta_{\max}$, and an oracle access to a nonexpansive map
		$$f: \calB\big((\alpha_{\min},\beta_{\min}),(\alpha_{\max},\beta_{\max})\big)\to \calB\big((\alpha_{\min},\beta_{\min}),(\alpha_{\max},\beta_{\max}),\pm \eps\big).$$
	}
	\KwOut{A point $x^*\in \calB\big((\alpha_{\min},\beta_{\min}),(\alpha_{\max},\beta_{\max})\big)$ such that $\norm{f(x^*)-x^*}_{\infty}\leq \eps$.
	}
	
	\BlankLine
	
	Let $\calA_1$ be an algorithm for $\NonExp^\dagger_{\infty}(\eps,a)$ and $\calA_2$ be an algorithm for $\NonExp^\dagger_{\infty}(\eps,b)$.\\

	\For{$t=1,2,\ldots$}{
	For each previous round $r<t$, let $\smash{q^r\in \calB(\beta_{\min},\beta_{\max})}$ be the point queried by $\calA_2$ (see line 4), let  $\smash{v^r\in \calB(\beta_{\min},\beta_{\max},\pm \eps)}$ be the answer to $q^r$ sent back to $\calA_2$ (see line~\ref{line: define v^t}),  and let $\smash{y^{(r,*)}}$ be the $\eps$-fixed point found by $\smash{\calA_1}$ during this round (see line~\ref{line: run A1}).\\

		Given  $((q^1,v^1),\cdots,(q^{t-1},v^{t-1}))$, let 
		  $q^t\in \calB(\beta_{\min},\beta_{\max})$ be the $t$-th query of $\calA_2$.\\
		
		
		Let (when $t=1$, set $p^1_{\min}\gets \alpha_{\min}$ and $p^1_{\max}\gets \alpha_{\max}$)\label{line: linfty construct}
		\begin{align*}p^t_{\min,i}&\leftarrow \max\left(\alpha_{\min,i},\max_{r\in[t-1]}\big(y^{(r,*)}_i-\norm{q^t-q^r}_{\infty}\big)\right) \text{ for all } i\in[a]\quad\text{and}\\[0.5ex]
		p^t_{\max,i}&\leftarrow \min\left(\alpha_{\max,i},\min_{r\in[t-1]}\big(y^{(r,*)}_i+\norm{q^t-q^r}_{\infty}\big)\right) \text{ for all } i\in[a].
		\end{align*}\label{define pr}\\
		
			Define a new function $g: \calB(p^t_{\min},p^t_{\max}) \to \calB(p^t_{\min},p^t_{\max},\pm \eps)$ as follows:
			$$ g(x)=\Big(f(x,q^t)_1,\cdots,f(x,q^t)_a\Big),
			\quad \text{for every $x\in \calB(p^t_{\min},p^t_{\max})$.}$$\\
			Run algorithm $\calA_1$ 
			  to find a solution $y^{(t,*)}\in\calB(p^t_{\min},p^t_{\max})$ to $\NonExp^\dagger_{\infty}(\eps,a)$ on $g$.\label{line: run A1}

		Set $\smash{v^t\in \calB(\beta_{\min},\beta_{\max},\pm \eps)}$ below as the answer to $q^t$ sent to $\calA_2$:\label{line: define v^t}
        $$v^t_{j-a}=f(y^{(t,*)},q^t)_j,\quad\text{for each $j\in[a+b]\setminus [a]$}.$$\\ 
		
		If $\norm{v^t-q^t}_{\infty}\leq \eps$, 
		  \Return $(y^{(t,*)},q^t)$.
	}
\end{algorithm}
\begin{proof}[Proof of \Cref{thm: linfty decomposition}]
	Let $\calA_1$ be an algorithm that solves $\NonExp^\dagger_{\infty}(\eps,a)$ using $q(\eps,a)$ queries and $t(\eps,a)$ time, and $\calA_2$ be an algorithm that solves $\NonExp^\dagger_{\infty}(\eps,b)$ using $q(\eps,b)$ queries and $t(\eps,b)$ time. We will show that \autoref{alg: linfty} can solve $\NonExp^\dagger_{\infty}(\eps,a+b)$ in $q(\eps,a)\cdot q(\eps,b)$ queries and $O((t(\eps,a)+q(\eps,b))\cdot q(\eps,b) + t(\eps,b))$ time.\medskip

\def\hhh{h}	
	
\noindent \textbf{Overview.}
Let $f$ the input nonexpansive map to the problem $\NonExp^\dagger_\infty(\eps,a+b)$, with $$f: \calB\big((\alpha_{\min},\beta_{\min}),(\alpha_{\max},\beta_{\max})\big)\to \calB\big((\alpha_{\min},\beta_{\min}),(\alpha_{\max},\beta_{\max}),\pm \eps\big),$$
where $\alpha_{\min}\preceq \alpha_{\max}\in [0,1]^a$ and 
$\beta_{\min}\preceq \beta_{\max}\in [0,1]^b$.
At a high level, \autoref{alg: linfty} 
  just runs $\calA_2$ for $\NonExp^\dagger_{\infty}(\eps,b)$ on a function $\hhh: \calB(\beta_{\min},\beta_{\max})$ $\to \calB(\beta_{\min},\beta_{\max},\pm \eps)$ built on the go as follows.
  
Let $q^1,\ldots,q^{t-1}\in \calB(\beta_{\min},\beta_{\max})$ be the $t-1$ queries that 
  $\calA_2$ has made on $h$ so far, for some $t\ge 1$, and 
  let $v^1,\ldots,v^{t-1}\in \calB(\beta_{\min},\beta_{\max},\pm \eps)$ be the query results
  of $q^1,\ldots,q^{t-1}$ on $\hhh$.
Let $q^t\in \calB(\beta_{\min},\beta_{\max})$ be the new query that $\calA_2$ makes in the $t$-th round.
We will use $f$ and the past queries and their associated results $((q^1,v^1),\cdots,(q^{t-1},v^{t-1}))$ to come up with a $v^t\in \calB(\beta_{\min},\beta_{\max},\pm \eps)$ 
  as the answer $h(q^t)$ to the query,
  such that the following two properties hold:
\begin{flushleft}\begin{enumerate} \item
\Cref{lemma: linfty consistent}: All results $(q^1,v^1),\ldots,(q^t,v^t)$ 
  are consistent with the nonexpansion property, i.e., $\norm{v^r-v^{r'}}_{\infty}\leq \norm{q^r-q^{r'}}_{\infty}$ for all $r,r'\in[t]$; and 
  \item \Cref{lemma: linfty find a solution}: When $q^t$ is a solution to $\NonExp^\dagger_{\infty}(\eps,b)$ on $h$,
  i.e., $\norm{v^t-q^t}_{\infty}\leq \eps$,
  we obtain a solution to $\NonExp^\dagger_{\infty}(\eps,a+b)$ with respect to 
  the original input function $f$.
\end{enumerate}\end{flushleft}

Note that the pairwise consistency of the answers (\Cref{lemma: linfty consistent}) implies that there exists a nonexpansive function $h$ such that $h(q^r) = v^r$ for all $r \in [t]$, by Lemma 11 of \cite{CLY25}.
As a result, $\calA_2$ must terminate within $q(\eps,b)$ many queries, with the last point queried being an $\eps$-fixed point of $h$.
It then follows from \Cref{lemma: linfty find a solution} that this $\eps$-fixed point of $h$ can be used to obtain an $\eps$-fixed point of the original function $f$.
So the main challenge is, given $q^t$ as the $t$-th query of $\calA_2$, what is the answer $v^t$ to give to $\calA_2$ so that both properties above (as in \Cref{lemma: linfty consistent} and \Cref{lemma: linfty find a solution}) hold.

To obtain the answer $v^t$ to the query $q^t$, intuitively, we fix the last $b$ coordinates to $q^t$ and run $\calA_1$ to find an $\eps$-fixed point of the projection of $f$ on the first $a$ coordinates.
The implementation, however, requires handling a number of intricate technical details.
As described on lines 5--8 of \autoref{alg: linfty}, we obtain $v^t$ by running $\calA_1$ on a function $g:\calB(p^t_{\min},p^t_{\max})\to \calB(p^t_{\min},p^t_{\max},\pm \eps)$, where 
\begin{flushleft}\begin{enumerate}
\item $p_{\min}^t,p_{\max}^t\in [0,1]^a$, satisfying $\alpha_{\min}\preceq p^t_{\min}\preceq p^t_{\max}\preceq \alpha_{\max}$, are carefully derived 
  from results of previous queries as well as the current query point $q^t$ to define the domain of $g$; and 
\item 
  $g$ is just the projection of $f$ on 
  $\calB(p^t_{\min},p^t_{\max})$ with the last $b$ coordinates fixed to be $q^t$.
\end{enumerate}\end{flushleft}
The most crucial idea here is to use the refined domain  $\calB(p^t_{\min},p^t_{\max})$ for $g$, rather than the full domain $\calB(\alpha_{\min},\alpha_{\max})$,
while showing that $g$ does not go outside of $\calB(p^t_{\min},p^t_{\max},\pm \eps)$. 
Because of this, $g$ is a valid input function to $\calA_1$, which must terminate with an $\eps$-fixed point of $g$, denoted by $y^{(t,*)}$.
The latter is then used to obtain $v^t$ on line 8.\medskip

	
\noindent	\textbf{Correctness.} 
We first prove \Cref{lemma: linfty consistent} and \Cref{lemma: linfty find a solution} which together, as discussed above, imply the correctness of   \autoref{alg: linfty} that it always returns an $\eps$-fixed point of $f$; we analyze the time and query complexities of \autoref{alg: linfty} at the end of this section.

We start by proving a lemma about    $p^t_{\min},p^t_{\max},g$ and $y^{(t,*)}$ during round $t$ of \autoref{alg: linfty}:

\begin{lemma}\label{hehehehe}
In each round $t\ge 1$, 
we have 
\begin{enumerate}
	\item $\alpha_{\min}\preceq p^t_{\min}\preceq p^t_{\max}\preceq \alpha_{\max}$;
	\item $g$ is a nonexpansive map with respect to the $\ell_{\infty}$-norm;
	\item $g(x)\in \calB(p^t_{\min},p^t_{\max},\pm \eps)$ for all $x\in \calB(p^t_{\min},p^t_{\max})$; and
	\item $\big\| f(y^{(t,*)},q^t)_{[a]}-y^{(t,*)} \big\|_{\infty}\leq \eps$.
\end{enumerate}
\end{lemma}

\begin{proof} 
We prove the lemma by induction on $t$. The base case when $t=1$ is trivial given that $p_{\min}^1 $ $=\alpha_{\min}$ and $p_{\max}^1=\alpha_{\max}$. For the induction step, we show that the four items above hold for $t>1$. We assume by the inductive hypothesis that  all four items hold for any $r\in[t-1]$.

Below we show each of the four items in order for round $t$.\medskip

\noindent\textbf{First item.} $\alpha_{\min}\preceq p_{\min}^t$ and $p_{\max}^t\preceq \alpha_{\max}$ follow directly from the definition. To show $p_{\min}^t\preceq p_{\max}^t$, it suffices to show that
\[y^{(r,*)}_i-\norm{q^t-q^r}_{\infty}\leq y^{(r',*)}_i+\norm{q^t-q^{r'}}_{\infty},\quad\text{for all $i\in[a]$ and $r,r'\in[t-1]$.}\]
Equivalently, we want to show that
$$\big\|y^{(r,*)}-y^{(r',*)}\big\|_{\infty}\leq \norm{q^t-q^{r'}}_{\infty}+\norm{q^t-q^r}_{\infty},\quad\text{for all $r<r'\in[t-1]$}.$$ Fix arbitrary $r<r'\in[t-1]$. By the inductive hypothesis, we have $p_{\min}^{r'}\preceq p_{\max}^{r'}$ from item 1 in round $r'$, and then  
$\smash{y^{(r',*)}\in \calB(p_{\min}^{r'},p_{\max}^{r'})}$ by combining items 1--3 in round $r'$. This implies that \[y^{(r,*)}_i-\norm{q^{r'}-q^r}_{\infty}\leq y^{(r',*)}_i\leq y^{(r,*)}_i+\norm{q^{r'}-q^r}_{\infty},\quad\text{for all $i\in[a]$}.\] So we know that $\norm{y^{(r,*)}-y^{(r',*)}}_{\infty}\leq \norm{q^{r'}-q^r}_{\infty}\leq \norm{q^{t}-q^r}_{\infty}+\norm{q^{t}-q^{r'}}_{\infty}.$

\vspace{0.2cm}
\noindent\textbf{Second item.} For the second item, notice that 
$$
\norm{g(x)-g(z)}_{\infty}\le \big\|{f(x,q^t)-f(z,q^t)}\big\|_{\infty}\leq \norm{x-z}_{\infty}.
$$
where the last inequality is because $f$ is a nonexpansive function.

\vspace{0.2cm}
\noindent\textbf{Third item.} For the third item, we first note an equivalent definiton of $\calB(p^t_{\min},p^t_{\max})$, i.e.,  
\begin{equation}\label{equation: Bpminpmax}
\calB(p^t_{\min},p^t_{\max})=\left(\bigcap_{r\in[t-1]}\calB\bigg(y^{(r,*)},y^{(r,*)},\pm \norm{q^t-q^r}_{\infty}\bigg)\right)\cap \calB(\alpha_{\min},\alpha_{\max}).
\end{equation}
Similarly, an equivalent definiton of $\calB(p^t_{\min},p^t_{\max},\pm \eps)$ is as follows:
\begin{equation}\label{equation: Bpminpmax eps}
\calB(p^t_{\min},p^t_{\max},\pm \eps)=\left(\bigcap_{r\in[t-1]}\calB\bigg(y^{(r,*)},y^{(r,*)},\pm \big(\norm{q^t-q^r}_{\infty}+\eps\big)\bigg)\right)\cap \calB(\alpha_{\min},\alpha_{\max},\pm \eps).
\end{equation}
Fix any $x\in \calB(p^t_{\min},p^t_{\max})$, then we show below that $g(x)\in \calB(p^t_{\min},p^t_{\max},\pm \eps)$.

We first note that $g(x)\in \calB(\alpha_{\min},\alpha_{\max},\pm \eps)$, which follows from the definition of $g$ and the fact that $f$ is a map from $\calB((\alpha_{\min},\beta_{\min}),(\alpha_{\max},\beta_{\max}))$ to $ \calB((\alpha_{\min},\beta_{\min}),(\alpha_{\max},\beta_{\max}),\pm \eps)$.

Next, fixing any $r\in[t-1]$, we will use $x\in \calB(y^{(r,*)},y^{(r,*)},\pm \norm{q^t-q^r}_{\infty})$ to show that 
$$
g(x)\in \calB\bigg(y^{(r,*)},y^{(r,*)},\pm \big(\norm{q^t-q^r}_{\infty}+\eps\big)\bigg),
$$
To this end, we note that
\begin{align*}
\norm{g(x)-y^{(r,*)}}_{\infty}&=\norm{f(x,q^t)_{[a]}-y^{(r,*)}}_{\infty}\\[0.2ex] &\leq \norm{f(x,q^t)_{[a]}-f(y^{(r,*)},q^r)_{[a]}}_{\infty}+\norm{f(y^{(r,*)},q^r)_{[a]}-y^{(r,*)}}_{\infty}.
\end{align*}
Since $f$ is nonexpansive, we know that $$\norm{f(x,q^t)_{[a]}-f(y^{(r,*)},q^r)_{[a]}}_{\infty}\leq \norm{(x,q^t)-(y^{(r,*)},q^r)}_{\infty}.$$
Since $x\in\calB(y^{(r,*)},y^{(r,*)},\pm \norm{q^t-q^r}_{\infty})$, we know that $$\norm{(x,q^t)-(y^{(r,*)},q^r)}_{\infty}\leq \norm{q^t-q^r}_{\infty}.$$
Moreover, we have $\norm{f(y^{(r,*)},q^r)_{[a]}-y^{(r,*)}}_{\infty}\leq \eps$ by the inductive hypothesis and thus,
$$
\norm{g(x)-y^{(r,*)}}_{\infty}\leq \norm{q^t-q^r}_{\infty}+\eps \quad\text{which implies}\quad g(x)\in \calB\bigg(y^{(r,*)},y^{(r,*)},\pm \big(\norm{q^t-q^r}_{\infty}+\eps\big)\bigg).
$$

\noindent\textbf{Fourth item.} Given that $g:\calB(p^t_{\min},p^t_{\max})\to \calB(p^t_{\min},p^t_{\max},\pm \eps)$ is nonexpansive, we know that the point $y^{(t,*)}$ that the algorithm $\calA_1$ returns must be an $\eps$-fixed point of $g$, i.e. $\norm{g(y^{(t,*)})-y^{(t,*)}}_{\infty}\leq \eps$, which implies from the definition of $g$ that $\norm{f(y^{(t,*)},q^t)_{[a]}-y^{(t,*)}}_{\infty}\leq \eps$.
\end{proof}

We are now ready to prove the two lemmas needed for the correctness of \autoref{alg: linfty}:

\begin{lemma}\label{lemma: linfty consistent}
	For every round $t\geq 1$ and every $r <t$, we have  $\norm{v^r-v^t}_{\infty}\leq \norm{q^r-q^t}_{\infty}$.
\end{lemma}
\begin{proof}
Since $v^t$ is defined by $f$ so clearly $v^t\in \calB(\beta_{\min},\beta_{\max},\pm \eps)$. Notice that
$$
\norm{v^t-v^r}_{\infty}=\max_{j\in[a+b]\setminus[a]} |f(y^{(t,*)},q^t)_j-f(y^{(r,*)},q^r)_j|\leq \norm{(y^{(t,*)},q^t) - (y^{(r,*)},q^r)}_{\infty}\leq \norm{q^t-q^r}_{\infty},
$$
where the last inequality follows from $y^{(t,*)}\in\calB(y^{(r,*)},y^{(r,*)},\pm \norm{q^t-q^r}_{\infty})$.
\end{proof}

\begin{lemma}\label{lemma: linfty find a solution}
	For every round $t$, if $\norm{v^t-q^t}_{\infty}\leq \eps$, then $\norm{f(y^{(t,*)},q^t)-(y^{(t,*)},q^t)}_{\infty}\leq \eps$.
\end{lemma}
\begin{proof}
	By the correctness of the algorithm $\calA_1$, we know that 
	$$
	\max_{j\in[a]} |f(y^{(t,*)},q^t)_j-(y^{(t,*)},q^t)_j|\leq \eps.
	$$
	If $\norm{v^t-q^t}_{\infty}\leq \eps$, then we know that 
	$$
	\max_{j\in[a+b]\setminus[a]} |f(y^{(t,*)},q^t)_j-(y^{(t,*)},q^t)_j|\leq \eps.
	$$
	Then the lemma follows.
\end{proof}

\noindent\textbf{Query complexity and time complexity.} For each round of \autoref{alg: linfty},  the algorithm $\calA_1$ is called once, which uses $q(\eps,a)$ queries and $t(\eps,a)$ time. The outer algorithm $\calA_2$ has no more than $q(\eps,b)$ rounds, which means the query complexity of \autoref{alg: linfty} is at most $q(\eps,a) \cdot q(\eps,b)$. In each round, line~\ref{line: linfty construct} takes $O(q(\eps,b))$ time, so each round of \autoref{alg: linfty} needs $O(q(\eps,b)+t(\eps,a))$ time, and the number of rounds is $q(\eps,b)$. The outer algorithm $\calA_2$ has running time $t(\eps,b)$. Overall the time complexity is $O((t(\eps,a)+q(\eps,b))\cdot q(\eps,b) + t(\eps,b))$.
\end{proof}	

\section{Implementation of the Query-Efficient Algorithm}\label{section: implementation}


\noindent In this section, we present an implementation of the query-efficient algorithm of \cite{CLY25}  with the following time complexity:

\theoremnontrivialimplementation*

\subsection{Balanced points and $\ell_{\infty}$-halfspaces}\label{section: balanced points and ell_infty halfspaces}

\noindent \textbf{Notation.} 
Given $x\in \mathbb{R}^k$ (not necessarily in $[0,1]^k$), an $i\in[k]$, and a sign $\phi\in\set{\pm 1}$, we write 
$$\calP_i(x,\phi):=\set{y\in[0,1]^k: \phi\cdot(y_i-x_i)= \norm{y-x}_{\infty}},$$ where $\calP$ is a shorthand for \emph{pyramid}. 
Given a 
  (Lebesgue) measurable set $S$, we use $\vol(S)$ to denote its Lebesgue volume or simply volume.

Given a measurable set $S\subseteq\bbR^k$ with $0<\vol(S)<\infty$, we say $x\in\bbR^k$ is a \emph{balanced point} of $S$ if
\[ \vol\big(\calP_i(x,-1)\cap S\big)=\vol\big(\calP_i(x,+1)\cap S\big),\quad\text{for every $i\in[k]$.} \]
More generally, for $\delta\ge 0$, we say $x$ is a \emph{$\delta$-balanced point} of $S$ if 
\[ \left|\vol\big(\calP_i(x,-1)\cap S\big)-\vol\big(\calP_i(x,+1)\cap S\big)\right| \leq \delta,
\quad\text{for every $i\in[k]$.}\]

We recall the following theorem in~\cite{CLY25}, which was the key to obtaining a query-efficient algorithm for contraction fixed points with respect to the $\ell_\infty$-norm. 

\begin{theorem}[\cite{CLY25}]
  \label{theorem: existence balanced point}
  Every measurable set $S\subseteq\bbR^k$ with $0<\emph{\vol}(S)<\infty$ has a balanced point.
\end{theorem}

The notion of a balanced point was introduced in~\cite{CLY25} in the study of query-efficient algorithms for $\ell_{\infty}$-contractions. Subsequently, in work on query-efficient algorithms for $\ell_p$-contractions with $p\in[1,\infty)\cup\set{\infty}$~\cite{haslebacher2025query}, the terms \emph{$\ell_p$-halfspace} and \emph{centerpoint} were proposed by analogy with the classical centerpoint theorem for halfspaces in Euclidean space. In the particular case of the $\ell_{\infty}$-norm, however, we retain the term \emph{balanced point}, as it better captures the geometric intuition arising from the pyramid decomposition. At the same time, we adopt the term \emph{$\ell_{\infty}$-halfspace} for the corresponding notion of a halfspace.

\begin{definition}[$\ell_{\infty}$-halfspace]\label{definition: ell_infty halfsapce}
Given $x\in\bbR^k$ and $v\in \mathbb{R}^k$, we call the following set $\calH_\infty(x,v)$ a closed $\ell_{\infty}$-halfspace with center $x$ and direction $v$: \[\calH_{\infty}(x,v)=\left(\bigcup_{i\in[k]:v_i\geq 0}\calP_i(x,+1)\right)\cup \left(\bigcup_{i\in[k]:v_i\leq 0}\calP_i(x,-1)\right).
\]
Note that when $v_i=0$, $\calH_\infty(x,v)$ contains both $\calP_i(x,+1)$ and $\calP_i(x,-1)$.
\end{definition}

With these definitions, it is easy to show that $x$ is a balanced point of $S$ if and only if for every $v\in\bbR^k$,  $\vol(\calH_{\infty}(x,v)\cap S)\ge \vol(S)/2.$ Moreover, if $x$ is a $\delta$-balanced point of $S$, then we have for every $v\in\bbR^k$,  $\vol(\calH_{\infty}(x,v)\cap S) \ge (\vol(S)-k\delta)/2.$



\subsection{The query-efficient algorithm of \cite{CLY25}}\label{sec:review}

We first note that to prove \Cref{theorem: non-trivial implementation}, it suffices to give an algorithm for $\Contraction_\infty(\eps,\gamma,k)$ with query complexity $O(k\log (1/\eps\gamma))$ and time complexity $(k\log (1/\eps\gamma))^{O(k)}$.
This is due to a~simple reduction from $\NonExp_{\infty}(\eps,k)$ to $\Contraction_{\infty}(\eps/2,\eps/2,k)$ with asymptotically the same query and time complexities given in~\cite{CLY25}. 
Focusing on $\Contraction_\infty(\eps,\gamma,k)$ from now on, 
 we first briefly review the $O(k\log(1/\eps\gamma))$-query algorithm for $\Contraction_{\infty}(\eps,\gamma,k)$ of~\cite{CLY25} (see also an explanation in~\cite{FH26}),
 and then show how to implement it in time $(k\log(1/\eps\gamma))^{O(k)}$. 

Given query access to an unknown $(1-\gamma)$-contraction map $f:[0,1]^k\rightarrow [0,1]^k$,
the algorithm works as follows. We initialize a candidate set $\Cand^0= [0,1]^k$ as the set of points that remain~possible to be the unique fixed point $x^*$ of $f$. In each round $t=1,2,\cdots$, we do the following steps:
\begin{itemize}
    \item Query \emph{some} $x$ and obtain $f(x)$;\vspace{-0.1cm}
    \item If $x$ is an $\eps$-fixed point of $f$, we are done; otherwise, update the candidate set as follows: $$\Cand^t\gets \Cand^{t-1}\setminus \calH_{\infty}(x, x-f(x)).$$
\end{itemize}
The updating rule above is justified by the following lemma from \cite{CLY25}:
\begin{lemma}[\cite{CLY25}]\label{lem:hehe1}
    If $x\in[0,1]^k$ is not an $\eps$-fixed point of $f$, then every $x'\in \calH_{\infty}(x, x-f(x))$ has $\norm{x'-x^*}_{\infty}> \eps\gamma/4$.
\end{lemma}

Given \Cref{lem:hehe1}, every point $x'$ with $\|x'-x^*\|_\infty\le \eps\gamma/4$ must remain in $\Cand^t$ and thus, the volume of the candidate set never drops below $\smash{({\eps\gamma}/4)^k}$.
So a query-efficient algorithm would look for a query point $x$ such that, regardless of $f(x)$, the volume of the candidate set (from $\Cand^{t-1}$ to $\Cand^t$) drops quickly. This is exactly where balanced points come in. 

For each round $t\geq 1$, the algorithm can query a balanced point of $\Cand^{t-1}$, which is guaranteed to exist by \Cref{theorem: existence balanced point}. Then we have that for each round $t\geq 0$, $|\Cand^t|\leq {1}/{2^t}$, which implies that the algorithm must have found an $\eps$-fixed point of $f$ within $O(k\log(1/\eps\gamma))$ queries.

This volume-reduction argument is reminiscent of the ellipsoid method, which uses central queries and linear cuts to construct successively smaller enclosing ellipsoids. Here, querying a balanced point instead removes an $\ell_\infty$-halfspace directly from the possibly nonconvex candidate set, reducing its volume by at least half at each unsuccessful query.

It can also be shown  that, even if the algorithm always queries a $\delta$-balanced point for sufficiently small $\delta$, e.g., \begin{equation}\label{eq:hehe2}\delta=\frac{1}{2k} \left(\frac{\eps\gamma}{4}\right)^k\end{equation} the algorithm still have query complexity at most  $O(k\log(1/\eps\gamma))$. (A formal argument for this can be found in the proof of \Cref{theorem: non-trivial implementation}.) 

As a result, to prove \Cref{theorem: non-trivial implementation}, it suffices to show how to compute a $\delta$-balanced point for the current candidate set with $\delta$ given in \Cref{eq:hehe2} in time $(k\log(1/\eps\gamma))^{O(k)}$, which we give next.

\subsection{A simple algorithm for balanced points}

Let $S\subseteq[0,1]^k$ be a measurable set with $\vol(S)>0$. In the proof of \Cref{theorem: existence balanced point} in~\cite{CLY25}, the following map $g_S$ over $[0,1]^k$ was defined: for any  $p\in[0,1]^k$,
\[
\big(g_S(p)\big)_i := p_i + \vol\big(\calP_i(p,+1)\cap S\big)-\vol\big(\calP_i(p,-1)\cap S\big).
\]
It was shown in~\cite{CLY25} that  $g_S$ is continuous and maps $[0,1]^k$ to itself. By Brouwer's fixed point theorem, $g_S$ has a fixed point, and it was shown that every fixed point of $g_S$ is a balanced point of $S$. The following new observation on $g_S$ is crucial for our    algorithm of finding a balanced point: 
\begin{lemma}\label{lemma: g is nonexpansive}
The map $g_S:[0,1]^k\rightarrow [0,1]^k$ is nonexpansive with respect to the $\ell_{\infty}$-norm.
\end{lemma}
\begin{proof}
We first recall a characterization of nonexpansive maps  with respect to the $\ell_{\infty}$-norm; here we present a high-dimensional version of Lemma 9 in~\cite{CLY25}:
\begin{claim}
    Let $h:[0,1]^k\rightarrow[0,1]^k$. If for any $x,y\in[0,1]^k$ such that $|x_i-y_i|=\norm{x-y}_{\infty}$ for every $i\in[k]$, we have $\norm{h(x)-h(y)}_{\infty}\leq \norm{x-y}_{\infty}$, then $h$ is nonexpansive with respect to the $\ell_{\infty}$-norm.
\end{claim}
\begin{proof}
Let $x\neq y\in[0,1]^k$ with $d:=\|x-y\|_\infty$.
For each $i\in[k]$, construct a piecewise-linear path
$\gamma_i:[0,d]\to[0,1]$ 
from $\gamma_i(0)=x_i$ to 
  $\gamma(d)=y_i$ such that the
  slope of each piece is
either $+1$ or $-1$. Such a path exists because
$
|x_i-y_i|\le d\le 1.
$
Indeed, assuming $x_i\le y_i$ without loss of generality, one may first move from $x_i$ to
\[
\frac{x_i+y_i+d}{2}
\]
with a slope $+1$ piece
and then back to $y_i$ with a slope $-1$ piece, provided this turning point is at most $1$.
Otherwise (when the turning point above is larger than $1$), one can first move from $x_i$ to
\[
\frac{x_i+y_i-d}{2}
\]
with a slope $-1$ piece
and then to $y_i$. Note that the second turning point is nonnegative because the sum of the two turning points is $d\le 1$ but we assumed that the first point is larger than $1$.

Let
$
\gamma(t):=(\gamma_1(t),\ldots,\gamma_k(t)).
$
Partition $[0,d]$ at all turning times of the coordinate paths:
\[
0=t_0<t_1<\cdots<t_m=d.
\]
On each interval $[t_{j-1},t_j]$, every coordinate of $\gamma$ has
constant slope of absolute value $1$. Hence
\[
\left|\gamma_i(t_j)-\gamma_i(t_{j-1})\right|
=t_j-t_{j-1}
=\left\|\gamma(t_j)-\gamma(t_{j-1})\right\|_\infty\quad \text{ for every } i.
\]
Thus the two endpoints are diagonal points, and the
assumption of the claim gives
\[
\left\|h(\gamma(t_j))-h(\gamma(t_{j-1}))\right\|_\infty
\le t_j-t_{j-1}.
\]
By the triangle inequality,
\[
\begin{aligned}
\|h(x)-h(y)\|_\infty
\le \sum_{j=1}^m
\left\|h(\gamma(t_j))-h(\gamma(t_{j-1}))\right\|_\infty \le \sum_{j=1}^m(t_j-t_{j-1})
=d
=\|x-y\|_\infty.
\end{aligned}
\]
This finishes the proof that $h$ is nonexpansive with respect to the
$\ell_\infty$-norm.
\end{proof}
Now let $p,q\in[0,1]^k$ be such that $|p_i-q_i|=d:=\norm{p-q}_{\infty}$ for every $i\in[k].$
Fix an $i\in [k]$. We show below that $|(g_S(q))_i-(g_S(p))_i|\le d$, from which the lemma follows.
To this end, let's assume without loss of generality that $q_i=p_i+d$ and it suffices to show that 
\begin{align}\label{eq:hehe5}
0\leq {}&
\vol\big(\calP_i(p,+1)\cap S\big)
-\vol\big(\calP_i(p,-1)\cap S\big)\\
&-\vol\big(\calP_i(q,+1)\cap S\big)
+\vol\big(\calP_i(q,-1)\cap S\big)
\leq 2d.\nonumber
\end{align} 

To prove this we start with some notation.
For every $z_{-i}\in[0,1]^{k-1}$, let
\[
r_p(z_{-i})=\max_{j\neq i}|z_j-p_j|
\quad\text{and}\quad
r_q(z_{-i})=\max_{j\neq i}|z_j-q_j|.
\]
Since $|p_j-q_j|=d$ for every $j$, we have
$
|r_p(z_{-i})-r_q(z_{-i})|\leq d.
$
For any $z\in [0,1]^k$, we have
\[
z=(z_i,z_{-i})\in\calP_i(p,+1)
\quad\Longleftrightarrow\quad
z_i\geq p_i+r_p(z_{-i}),
\]
and
\[
z=(z_i,z_{-i})\in\calP_i(p,-1)
\quad\Longleftrightarrow\quad
z_i\leq p_i-r_p(z_{-i}),
\]
and similarly for $q$. Moreover, we have
\[
0\leq
\big(q_i+r_q(z_{-i})\big)
-\big(p_i+r_p(z_{-i})\big)
\leq 2d
\quad\text{and}\quad
0\leq
\big(q_i-r_q(z_{-i})\big)
-\big(p_i-r_p(z_{-i})\big)
\leq 2d,
\]
and the sum of the two differences above is exactly $2d$.

Now we prove \Cref{eq:hehe5}.
We start with the left inequality.
First, $q_i =p_i +d$ and $|r_p(z_{-i})-r_q(z_{-i})| \leq d$ imply that if $z_i \ge q_i+r_q(z_{-i})$ then also $z_i \ge p_i+r_p(z_{-i})$; thus, if $z \in \calP_i(q,+1)$ then also $z \in \calP_i(p,+1)$, i.e., $\calP_i(q,+1) \subseteq \calP_i(p,+1)$. Hence also
$\calP_i(q,+1) \cap S \subseteq \calP_i(p,+1) \cap S$.
Similarly, $\calP_i(p,-1) \subseteq \calP_i(q,-1)$ and $\calP_i(p,-1) \cap S\subseteq \calP_i(q,-1) \cap S$.
These imply the left inequality.

For the right inequality, note that $z\in \calP_i(p,+1) \setminus \calP_i(q,+1)$ iff $q_i +r_q(z_{-i}) > z_i \ge p_i+r_p(z_{-i})$, i.e. iff $z_i$ lies in an interval of length $(q_i +r_q(z_{-i})) - (p_i+r_p(z_{-i}))$. Similarly, $z \in\calP_i(q,-1) \setminus \calP_i(p,-1)$ iff $(p_i-r_p(z_{-i})) < z_i \le (q_i-r_q(z_{-i}))$, i.e. iff $z_i$ lies in an interval of length $(q_i -r_q(z_{-i})) - (p_i-r_p(z_{-i})) $. The sum of the two lengths is $2d$. Since all coordinates are in $[0,1]$, it follows by integrating
over $z_i$ and $z_{-i}$ that $\vol(\calP_i(p,+1)\cap S)
-\vol(\calP_i(p,-1)\cap S)-
\vol(\calP_i(q,+1)\cap S)
+\vol(\calP_i(q,-1)\cap S)
\leq 2d$.
This finishes the proof of \Cref{eq:hehe5} 
and thus, \Cref{lemma: g is nonexpansive}.
\end{proof}
Next we record two
algorithmic ingredients needed in our algorithm. The first follows
from the fixed-point algorithm of Shellman and Sikorski~\cite{SS03}:

\begin{lemma}[\cite{SS03}]\label{lemma:ss-algorithm}
Let $h:[0,1]^k\rightarrow[0,1]^k$ be nonexpansive under the
$\ell_{\infty}$-norm. Suppose that each evaluation of $h$ takes time $T$.
Then, for every $\delta>0$, a $\delta$-fixed point of $h$ can be computed
in time
\[
O\left(\log^k(1/\delta)\cdot T\right).
\]
\end{lemma}

We also use the following volume-computation result of Feodorov and
Haslebacher~\cite{FH26}: 
\begin{lemma}[\cite{FH26}]
\label{lemma:volume-computation}
Let $S$ be a set obtained  by removing $t$ many $\ell_{\infty}$-halfspaces from $[0,1]^k$, where centers and directions of the $\ell_\infty$-halfspaces are given explicitly.
Then given any $p\in[0,1]^k$, $i\in[k]$, and $\phi\in\set{\pm1}$, the volume $\emph{\vol}(\calP_i(p,\phi)\cap S)$
can be computed in time $(tk)^{O(k)}$.
\end{lemma}

We are now ready to prove
\Cref{theorem: non-trivial implementation}.

\begin{proof}[Proof of \Cref{theorem: non-trivial implementation}]
Fix $\delta$ to be the value given in 
  \Cref{eq:hehe2}.

Consider round $t$ of the query algorithm, and let $S\coloneqq\Cand^{t-1}$ be the current candidate set. By
\Cref{lemma: g is nonexpansive}, the map $g_S$ is nonexpansive with
respect to the $\ell_{\infty}$-norm. Hence, by
\Cref{lemma:ss-algorithm}, we can compute a point $p\in[0,1]^k$
satisfying $\norm{g_S(p)-p}_{\infty}\leq\delta.$
For every $i\in[k]$, the definition of $g_S$ gives
\[
\big(g_S(p)\big)_i-p_i
=
\vol\big(\calP_i(p,+1)\cap S\big)
-
\vol\big(\calP_i(p,-1)\cap S\big).
\]
It follows that
\[
\left|
\vol\big(\calP_i(p,-1)\cap S\big)
-
\vol\big(\calP_i(p,+1)\cap S\big)
\right|
\leq\delta
\]
for every $i\in[k]$. Thus, $p$ is a $\delta$-balanced point of $S$.

We query $f$ at $p$. If $p$ is an $\eps$-fixed point, the algorithm
terminates. Otherwise, we update
\[
\Cand^t
=
\Cand^{t-1}\setminus\calH_{\infty}(p,p-f(p)).
\]
By \Cref{lem:hehe1}, the set
\[
\left\{
x\in[0,1]^k:
\norm{x-x^*}_{\infty}\leq \frac{\eps\gamma}{4}
\right\}
\]
remains in the candidate set throughout the algorithm. So we have
\begin{equation}\label{eq:hehe3}
\vol(S)=\vol(\Cand^{t-1})\geq \left(\frac{\eps\gamma}{4}\right)^k.
\end{equation}
Moreover, since $p$ is $\delta$-balanced,
\[
\vol\bigg(
S\cap\calH_{\infty}\big(p,p-f(p)\big)
\bigg)
\geq
\frac{\vol(S)-k\delta}{2}\ \ \Longrightarrow\ \ 
\vol(\Cand^t)
\leq
\frac{\vol(S)+k\delta}{2}
\le \frac{3}{4}\cdot \vol(\Cand^{t-1})
\]
where the last inequality used \Cref{eq:hehe3} and the choice of $\delta$
  in \Cref{eq:hehe2}.
Consequently, the algorithm~terminates within $O\left(k\log(1/\eps\gamma)\right)$
rounds and makes the same number of queries to $f$.

It remains to bound the running time. At the beginning of round $t$,
$S:=\Cand^{t-1}$ is obtained from $[0,1]^k$ by removing
$t-1$ many $\ell_{\infty}$-halfspaces. 
By \Cref{lemma:volume-computation}, each evaluation of $g_S$ can be done in time $(tk)^{O(k)}.$
Combining with \Cref{lemma:ss-algorithm}, the query point
in round $t$ can be computed in time
\[
O\left(\log^k(1/\delta)\cdot (tk)^{O(k)}\right).
\]
Plugging in $t=O(k\log(1/\eps\gamma))$,
\[
\log\frac{1}{\delta}
=
k\log\frac{4}{\eps\gamma}+\log(2k)
=
O\left(k\log\frac{1}{\eps\gamma}\right),
\]
and multiplying by the number of rounds $O(k\log(1/\eps\gamma))$ give the desired overall running time of 
\[
\left(k\log\frac{1}{\eps\gamma}\right)^{O(k)}.
\]

This gives the claimed query and time complexities for
$\NonExp_{\infty}(\eps,k)$ after applying the reduction from $\NonExp_\infty(\eps,k)$ to $\Contraction_\infty(\eps/2,\eps/2,k)$ given in \cite{CLY25}.
\end{proof}

\section*{Acknowledgment and AI disclosure}
Xi Chen and Yuhao Li were supported by NSF grants CCF-2106429 and CCF-2107187, and a grant from the AI Security Institute.  Mihalis Yannakakis was supported by NSF grants CCF-2107187, CCF-2212233, and CCF-2332922.

The authors had worked out both results by February 2026. AI tools were only used to assist with language editing.

\begin{flushleft}
\bibliographystyle{alpha}
\bibliography{ref}
\end{flushleft}

\end{document}